# Effect of $Z'$-mediated flavor-changing neutral current on $B \to \pi\pi$ decays

**S. Sahoo[1*], C. K. Das[2], and L. Maharana[3]**

[1]Department of Physics, National Institute of Technology, Durgapur, West Bengal, India
[2]Department of Physics, Trident Academy of Technology, Bhubaneswar, Orissa, India.
[3]Department of Physics, Utkal University, Bhubaneswar, Orissa, India.
[*]E-mail: sukadevsahoo@yahoo.com

**ABSTRACT**

We study the effect of $Z'$-mediated flavor-changing neutral current on the $B \to \pi\pi$ decays. The branching ratios of these decays can be enhanced remarkably in the non-universal $Z'$ model. Our estimated branching ratios $\mathrm{B}(B^0 \to \pi^0\pi^0)$ are enhanced significantly from their standard model (SM) value. For $g'/g = 1$, the branching ratios $\mathrm{B}(B^0 \to \pi^0\pi^0)$ are very close to the recently observed experimental values and for higher values of $g'/g$ branching ratios are more. Our calculated branching ratios $\mathrm{B}(B^0 \to \pi^+\pi^-)$ and $\mathrm{B}(B^+ \to \pi^+\pi^0)$ are also enhanced from the SM value as well as the recently observed experimental values. These enhancements of branching ratios from their SM value give the possibility of new physics.

## 1. INTRODUCTION

The decays of $B$ mesons [1–5] provide information about the flavor structure of the standard model (SM), the origin of CP violation, the dynamics of hadronic decays, and to search for any signals of new physics beyond the SM. The $B$ factories, such as Belle (KEK) [6,7] and BaBar (SLAC) [8,9] have provided us huge data in this direction. The main objective of these $B$ factories is to critically test the standard model predictions and to look for possible signatures of new physics (NP). The $B \to \pi\pi$ system offers three decay channels, $B^0 \to \pi^0\pi^0$, $B^0 \to \pi^+\pi^-$ and $B^+ \to \pi^+\pi^0$, as well as their CP conjugates [10–13]. Recent experimental measurements [5–9] of CP asymmetries and the branching ratios of $B \to \pi\pi$ decays have shown deviations from the SM expectations. In the $B \to \pi\pi$ decays, there are three discrepancies: (i) the direct CP asymmetry for the mode $B^0 \to \pi^+\pi^-$ is very large; (ii) $B^0 \to \pi^0\pi^0$ mode is found to have larger branching ratio $(\approx 1.3 \times 10^{-6})$ [5] than the SM expectations $(\approx 10^{-7})$ [4]; and (iii) the theoretical estimation of $B^0 \to \pi^+\pi^-$ branching ratio is about 2 times larger than the current experimental average. This feature is the "$B \to \pi\pi$ Puzzle". The large branching ratio of the $B^0 \to \pi^0\pi^0$ channel tells us that if unknown electroweak penguin (EWP) dynamics is the cause of the enhancement, the EWP amplitude should be quite large. It is difficult to have such a large amplitude from intuitive arguments. That is why we prefer a new physics (NP) solution for the puzzle.

There are at least *three* kinds of models [14] used for theoretical calculations of the branching ratios and CP asymmetries ($A_{CP}$) for nonleptonic charmless *B* decay modes: the conventional factorization (CF) model [15], the QCD-improved factorization (QCDF) model [16] and the perturbative QCD (PQCD) model [17]. The main problem is to calculate the strong phase and hence the CP asymmetry theoretically. In CF model the strong phases are calculated from the imaginary parts of the respective Wilson coefficients. QCDF predicts a small strong phase difference between the dominant amplitudes. In PQCD method hard components, which are treated by perturbation theory, are separated from a QCD process. Nonperturbative components can be extracted from experimental data. In PQCD predictions, annihilation and exchange topologies are given more weight than in QCDF. They can generate a sizable strong phase. $B \to \pi\pi$ and related decays are also studied in the heavy quark limit of QCD using the soft collinear effective theory (SCET) [18]. QCDF and SCET are based on collinear factorization theorem, but PQCD is based on $k_T$ factorization theorem.

The *B* meson decays [19] induced by the flavor-changing neutral current (FCNC) transitions are very important to probe the quark-flavor sector of the SM. In the SM they arise from one-loop diagrams and are generally suppressed in comparison to the tree diagrams. Nevertheless, one-loop FCNC processes can be enhanced by orders of magnitude in some cases due to the presence of new physics. New physics comes into play in *B* meson decays in *two* different ways: (a) through a new contribution to the Wilson coefficients, and (b) through a new structure in the effective Hamiltonian, which are both absent in the SM. In this paper, we study $B \to \pi\pi$ decays considering the effect of $Z'$-mediated FCNC which modifies the Wilson coefficients and changes the effective Hamiltonian, and gives a new result for the branching ratio.

In the $Z'$ sector, there has been a great deal of investigation to understand the underlying physics beyond the SM [20–28]. With flavor-changing neutral currents [25,29–31], the $Z'$ boson contributes at tree level, and its contribution will interfere with the SM contributions. The $Z'$ flavor-changing couplings will give new contributions to the SM operators. The new contributions from flavor-changing $Z'$ models in which $Z'$ mediates vector and axial vector interactions would enhance the branching ratios of $B \to \pi\pi$ decays significantly [32].

This paper is organized as follows: in Section 2, we give a brief account of $B \to \pi\pi$ decays in the standard model and then discuss the "$B \to \pi\pi$ puzzle". In Section 3, we evaluate the effective Hamiltonian and corresponding branching ratios for $B \to \pi\pi$ decays considering the contribution from $Z'$ boson. In Section 4, we discuss the results, so obtained, and compare our results with that of the standard model as well as the recently observed experimental values.

## 2. $B \to \pi\pi$ DECAYS IN THE STANDARD MODEL

Let us consider the $B \to \pi\pi$ decay processes. In the standard model, these decays involve $\bar{b} \to \bar{d} q \bar{q}$ (q = u,d) transitions through exchange of W-boson. In the $B \to \pi\pi$ decays [33], the B meson is heavy, sitting at rest. It decays into two light mesons with large momenta. Therefore the light mesons are moving very fast in the rest frame of B meson. In this case, the short distance hard process dominates the decay amplitude. The effective Hamiltonian [34–37] describing the $B \to \pi\pi$ decays can be written as:

$$H_{\text{eff}}^{\text{SM}} = \frac{G_F}{\sqrt{2}} \sum_{q=u,c} \lambda_q \left[ C_1 Q_1^q + C_2 Q_2^q + \sum_{i=3}^{10} C_i Q_i + C_g Q_g \right] + h.c., \quad (1)$$

where $\lambda_q = V_{qb}^* V_{qd}$, $C_i$'s are Wilson coefficients and $Q_i$'s are local operators containing quark and gluon fields [34]. Both the $C_i$'s and $Q_i$'s depend on the QCD renormalization scale $\mu$ i.e. $C_1(\mu)$, $Q_1(\mu)$ etc. and $C_i$'s depend on the mass of the W boson and the masses of other heavy particles such as the top quark as well. The renormalization scale $\mu$ is typically of the order of a few GeV. The current-current operators induced by W-boson exchange are given by:

$$Q_1^q = \left(\bar{b}_\alpha q_\beta\right)_{V-A} \left(\bar{q}_\beta d_\alpha\right)_{V-A}, \quad Q_2^q = \left(\bar{b}_\alpha q_\alpha\right)_{V-A} \left(\bar{q}_\beta d_\beta\right)_{V-A}. \quad (2)$$

The $Q_i$'s, $i = 3, 4, 5, 6$ i.e.

$$Q_3 = \left(\bar{b}_\alpha d_\alpha\right)_{V-A} \sum_q \left(\bar{q}_\beta q_\beta\right)_{V-A}, \quad (3)$$

$$Q_4 = \left(\bar{b}_\alpha d_\beta\right)_{V-A} \sum_q \left(\bar{q}_\beta q_\alpha\right)_{V-A},$$

$$Q_5 = \left(\bar{b}_\alpha d_\alpha\right)_{V-A} \sum_q \left(\bar{q}_\beta q_\beta\right)_{V+A},$$

$$Q_6 = \left(\bar{b}_\alpha d_\beta\right)_{V-A} \sum_q \left(\bar{q}_\beta q_\alpha\right)_{V+A},$$

are the QCD penguin operators, summed over the quark flavors $q = u, d, s, c, b$ and the other $Q_i$'s, $i = 7, 8, 9, 10$ i.e.

$$Q_7 = \frac{3}{2} \left(\bar{b}_\alpha d_\alpha\right)_{V-A} \sum_q e_q \left(\bar{q}_\beta q_\beta\right)_{V+A}, \quad (4)$$

$$Q_8 = \frac{3}{2} \left(\bar{b}_\alpha d_\beta\right)_{V-A} \sum_q e_q \left(\bar{q}_\beta q_\alpha\right)_{V+A},$$

$$Q_9 = \frac{3}{2} \left(\bar{b}_\alpha d_\alpha\right)_{V-A} \sum_q e_q \left(\bar{q}_\beta q_\beta\right)_{V-A},$$

$$Q_{10} = \frac{3}{2}(\bar{b}_\alpha d_\beta)_{V-A} \sum_q e_q (\bar{q}_\beta q_\alpha)_{V-A} ,$$

are the electroweak penguin operators, where $e_q$ is the charge of the quark, and lastly

$$Q_g = \frac{g_s}{8\pi^2} m_b \bar{b}\sigma^{\mu\nu}(1-\gamma_5)G_{\mu\nu} d , \qquad (5)$$

is the chromomagnetic dipole operator. Here, $(\bar{q}_1 q_2)_{V \pm A} = \bar{q}_1 \gamma^\mu (1 \pm \gamma_5) q_2$.

The topological amplitudes provide a parametrization for non-leptonic *B*-meson decay processes which is independent of theoretical models for the calculation of hadronic matrix elements [38]. The decay amplitudes of $B \to \pi\pi$ [17] can be written as:

$$A(B^0 \to \pi^+\pi^-) = -T\left(1 + \frac{P}{T} e^{i\phi_2}\right), \qquad (6)$$

$$\sqrt{2} A(B^0 \to \pi^0\pi^0) = T\left[\left(\frac{P}{T} - \frac{P_{EW}}{T}\right)e^{i\phi_2} - \frac{C}{T}\right],$$

$$\sqrt{2} A(B^+ \to \pi^+\pi^0) = -T\left[1 + \frac{C}{T} + \frac{P_{EW}}{T} e^{i\phi_2}\right],$$

where *T*, *C*, *P* and $P_{EW}$ stand for the color-allowed tree, color-suppressed tree, penguin, and electroweak penguin amplitudes, respectively, and $\phi_2$ is the weak phase defined by $V_{ub} = |V_{ub}|\exp(-i\phi_3)$, $V_{td} = |V_{td}|\exp(-i\phi_1)$ and $\phi_2 = 180^0 - \phi_1 - \phi_3$.

The branching ratio corresponding to the decay amplitude can be written as

$$B(B \to \pi\pi)|_{SM} = \frac{1}{8\pi} \frac{|\mathbf{P}|}{m_B^2} |A(B \to \pi\pi)|^2 \frac{1}{\Gamma_{tot}} , \qquad (7)$$

where $|\mathbf{P}|$ is the 3-momentum of the final state particles in the rest frame of the *B*-meson [18,39]. The theoretical predictions for the *CP*-averaged branching ratios of the decay modes of $B \to \pi\pi$ are given by Benke and Neubert [40] as:

$$B(B^0 \to \pi^+\pi^-) = 8.9^{+4.0}_{-3.4} \times 10^{-6}, \qquad (8)$$

$$B(B^0 \to \pi^0\pi^0) = 0.3^{+0.2}_{-0.2} \times 10^{-6},$$

$$B(B^+ \to \pi^+\pi^0) = 6.0^{+3.0}_{-2.4} \times 10^{-6}.$$

The recently observed branching ratios of $B \to \pi\pi$ decays [5] (Updated August 2006) are given as:

$$\mathrm{B}\left(B^0 \to \pi^+\pi^-\right) = (5.2 \pm 0.2) \times 10^{-6}, \quad (9)$$

$$\mathrm{B}\left(B^0 \to \pi^0\pi^0\right) = (1.31 \pm 0.21) \times 10^{-6},$$

$$\mathrm{B}\left(B^+ \to \pi^+\pi^0\right) = (5.7 \pm 0.4) \times 10^{-6}.$$

These measurements represent a challenge for theoretical study. For example, the theoretical estimation of $B^0 \to \pi^+\pi^-$ branching ratio is about 2 times larger than the current experimental average. On the other hand, the calculation of $B^+ \to \pi^+\pi^0$ reproduces the data rather well. This "$B \to \pi\pi$ Puzzle" is reflected by the following quantities [2,41]:

$$R_{00}^{\pi\pi} = 2\left[\frac{\mathrm{B}\left(B^0 \to \pi^0\pi^0\right) + \mathrm{B}\left(\bar{B}^0 \to \pi^0\pi^0\right)}{\mathrm{B}\left(B^0 \to \pi^+\pi^-\right) + \mathrm{B}\left(\bar{B}^0 \to \pi^+\pi^-\right)}\right] = (0.66 \pm 0.14)_{\exp}, \quad (10)$$

$$R_{+-}^{\pi\pi} = 2\left[\frac{\mathrm{B}\left(B^+ \to \pi^+\pi^0\right) + \mathrm{B}\left(B^- \to \pi^-\pi^0\right)}{\mathrm{B}\left(B^0 \to \pi^+\pi^-\right) + \mathrm{B}\left(\bar{B}^0 \to \pi^+\pi^-\right)}\right]\frac{\tau_{B^0}}{\tau_{B^+}} = (2.30 \pm 0.35)_{\exp},$$

where $\tau_{B^+}/\tau_{B^0} = 1.069$, the central values calculated within the QCD factorization give $R_{+-}^{\pi\pi} = 1.24$ and $R_{00}^{\pi\pi} = 0.07$ [40]. These numbers show the discrepancies between theoretical and experimental predictions.

The counting rules in terms of powers of the Wolfenstein parameter $\lambda \sim 0.22$ [42] are assigned to various decay amplitudes [38,43]. In $B \to \pi\pi$ decays, the dominant contribution comes from T. The amplitudes in Eq. (6), obey the counting rules in the standard model [38],

$$\frac{P}{T} \sim \lambda, \quad \frac{C}{T} \sim \lambda, \quad \frac{P_{EW}}{T} \sim \lambda^2. \quad (11)$$

Here, the use of parameter $\lambda$ is not related to CKM matrix elements. It is simply used as a measure of the relative sizes of various contributions. For example, $|C/T| \sim \lambda$ is due to color suppression. The hierarchy of the branching ratio $\mathrm{B}\left(B^0 \to \pi^0\pi^0\right) \sim O(\lambda^2)$ $\mathrm{B}\left(B^0 \to \pi^+\pi^-\right)$ is then expected. But the experimental results given in equation (9) show that the former is about of $O(\lambda)$ of the latter. This is the "$B \to \pi\pi$ Puzzle". The electroweak penguin (EWP) effects will be more important when there is a non-strange neutral particle in the final state, such as $\pi^0, \eta, \rho^0$ or $\phi$, as the color-allowed electroweak penguin $P_{EW}$ is involved. All charged final state will be less affected by the presence of electroweak penguin diagram since in this case only color-suppressed electroweak penguin $P_{EW}^C$ diagram can arise.

It has been claimed that the "$B \to \pi\pi$ puzzle" is resolved in the QCD-improved factorization (QCDF) approach [16] with an input from soft-collinear effective theory (SCET) [40,44]: the inclusion of NLO jet function, the hard coefficient of SCET$_{\mathrm{II}}$, into the QCDF formula for the color-suppressed tree amplitude gives sufficient enhancement of the $B^0 \to \pi^0\pi^0$ branching ratio. But the color-suppressed tree amplitude cannot be explained. Final-state interaction (FSI) is a plausible resolution to the "$B \to \pi\pi$ puzzle"; but the estimate of its effect is quite model-dependent. It is found that FSI produces a better agreement between theory and experiment in the measurement of the branching ratios of $B^0 \to \pi^0\pi^0$ and $B^0 \to \pi^+\pi^-$ decay modes in the Regge model [45]. Different methods for detection and measurement of new physics in $B \to \pi\pi$ decays have been studied in [46]. It has been shown that many models beyond the standard model could enhance the branching ratio and consequently tried to resolve the puzzle. Recently, Yang et al. [41] study $B \to \pi\pi$ decays in the minimal supersymmetric standard model with $R$-parity violation. They showed that R-parity violation can resolve the discrepancies in $B \to \pi\pi$ decays.

## 3. EFFECT OF $Z'$ BOSON ON $B \to \pi\pi$ DECAYS

Many models beyond the SM predict the existence of exotic fermions. These new (exotic) fermions can mix with the SM fermions. Such mixing induces FCNCs [25]. If these exotic fermions have different $U(1)'$ charges from the ordinary fermions, as found in $E_6$ models [29], interesting phenomena arise. Mixing between ordinary (doublet) and exotic singlet left-handed fermions induces undesirable FCNC, mediated by the SM Z boson. The *mixing* of the right-handed ordinary and exotic fermions induces FCNC mediated by $Z'$ boson.

In this paper, we consider the models in which the interactions between the $Z'$ boson and fermions are flavor nonuniversal for left-handed couplings and flavor diagonal for right-handed couplings. The basic formalism of the family non-universal $Z'$ models with flavor-changing neutral currents can be found in [25,47,48], to which we refer readers for detail. The family non-universal $Z'$ couplings lead to flavor-changing (non-diagonal) $Z'$ couplings and possibly to new effects, when quark and lepton flavor mixing are taken account.

In these models, the $Z'$ part of the neutral-current Lagrangian in the gauge basis can be written as

$$L^{Z'} = -g' J'_\mu Z'^\mu, \qquad (12)$$

where $g'$ is the gauge coupling constant of the $U(1)'$ group at the $M_W$ scale. The $Z'$ chiral current is

$$J'_\mu = \sum_{i,j} \overline{\psi}^I_i \gamma_\mu \left[ \left(\epsilon_{\psi L}\right)_{ij} P_L + \left(\epsilon_{\psi R}\right)_{ij} P_R \right] \psi^I_j, \qquad (13)$$

where the sum extends over all flavors of the SM fermion fields, $P_{L,R} \equiv (1 \mp \gamma_5)/2$ are the chirality projection operators, the superscript I refers to the gauge interaction eigenstates, and $\epsilon_{\psi L}$ ($\epsilon_{\psi R}$) denotes the left-handed (right-handed) chiral coupling matrix. The mass eigenstates of the chiral fields are $\psi_{L,R} = V_{\psi L,R} \psi^I_{L,R}$ and the CKM matrix is given by $V_{CKM} = V_{uL} V^\dagger_{dL}$. The chiral $Z'$ coupling matrices in the physical basis of up-type and down-type are,

$$B^X_u = V_{uX} \in_{uX} V^\dagger_{uX}, \qquad B^X_d = V_{dX} \in_{dX} V^\dagger_{dX}, \quad (X = L, R) \qquad (14)$$

where $B^X_{u(d)}$ are hermitian. As long as the $\in$ matrices are not proportional to the identity, the B matrices will have non-zero off-diagonal elements that induce FCNC interactions at tree level.

The effective Hamiltonian of the $\bar{b} \to \bar{p} q \bar{q}$ (p = d, s and q = u, d, s) transitions mediated by the $Z'$ boson can be written as [47–49]:

$$H^{Z'}_{eff} = \frac{2G_F}{\sqrt{2}} \left( \frac{g' M_Z}{g M_{Z'}} \right)^2 B^{L*}_{pb} (\bar{p}b)_{V-A} \sum_q \left[ B^L_{qq} (\bar{q}q)_{V-A} + B^R_{qq} (\bar{q}q)_{V+A} \right] + h.c., \qquad (15)$$

where $g = e/(\sin\theta_W \cos\theta_W)$ and $g'$ is the gauge coupling associated with the $U(1)'$ group. $B^L_{ij}$ and $B^R_{ij}$ refer to the left- and right-handed effective $Z'$ couplings of the quarks $i$ and $j$ at the weak scale respectively. The forms of the above four-quark operators $(\bar{p}b)_{V-A}(\bar{q}q)_{V-A}$ and $(\bar{p}b)_{V-A}(\bar{q}q)_{V+A}$ already exist in the SM. With FCNCs, the $Z'$ boson contributes at tree level, and its contribution will interfere with the SM contributions. The $Z'$ boson contributes to the QCD penguin operators $Q_{3(5)}$ as well as electroweak penguin operators $Q_{7(9)}$. The Wilson coefficients of the corresponding operators are modified due to $Z'$ effect. The effective Hamiltonian given by equation (15) can be written as:

$$H^{Z'}_{eff} = -\frac{G_F}{\sqrt{2}} V_{tb} V^*_{tp} \sum_q \left( \Delta C_3 Q^q_3 + \Delta C_5 Q^q_5 + \Delta C_7 Q^q_7 + \Delta C_9 Q^q_9 \right) + h.c., \qquad (16)$$

where $\Delta C_i$ denote the modifications to the corresponding SM Wilson coefficients induced by the $Z'$ gauge boson, which can be expressed as:

$$\Delta C_{3,5} = -\frac{2}{3 V_{tb} V^*_{tp}} \left( \frac{g' M_Z}{g M_{Z'}} \right)^2 B^L_{pb} \left( B^{L,R}_{uu} + 2 B^{L,R}_{dd} \right), \qquad (17)$$

$$\Delta C_{9,7} = -\frac{4}{3 V_{tb} V^*_{tp}} \left( \frac{g' M_Z}{g M_{Z'}} \right)^2 B^L_{pb} \left( B^{L,R}_{uu} - B^{L,R}_{dd} \right).$$

Generally, the diagonal elements of the effective coupling matrices $B_{qq}^{L,R}$ are real due to the hermicity of the effective Hamiltonian, but the off-diagonal elements $B_{pb}^{L}$ may contain a new weak phase $\phi_p^L$. Then the $\Delta C_i$'s can be represented as [49]:

$$\Delta C_{3,5} = 2 \frac{|V_{tb} V_{tp}^*|}{V_{tb} V_{tp}^*} \zeta_p^{LL,LR} e^{i\phi_p^L} \quad , \quad \Delta C_{9,7} = 4 \frac{|V_{tb} V_{tp}^*|}{V_{tb} V_{tp}^*} \xi_p^{LL,LR} e^{i\phi_p^L} . \qquad (18)$$

where the newly introduced $Z'$ parameters $\zeta_p^{LL,LR}$, $\xi_p^{LL,LR}$ and $\phi_p^L$ are defined as

$$\zeta_p^{LL,LR} = -\frac{1}{3}\left(\frac{g' M_Z}{g M_{Z'}}\right)^2 \frac{|B_{pb}^L|}{V_{tb} V_{tp}^*}\left(B_{uu}^{L,R} + 2 B_{dd}^{L,R}\right), \qquad (19)$$

$$\xi_p^{LL,LR} = -\frac{1}{3}\left(\frac{g' M_Z}{g M_{Z'}}\right)^2 \frac{|B_{pb}^L|}{V_{tb} V_{tp}^*}\left(B_{uu}^{L,R} - B_{dd}^{L,R}\right),$$

$$\phi_p^L = \arg[B_{pb}^L].$$

It is also noted that the other SM Wilson coefficients may receive contributions from the $Z'$ boson through renomalization group (RG) evolution. We assume that there is no significant RG running effect between the $M_W$ and $M_{Z'}$ scales. Hence, the RG evolution of the modified Wilson coefficients is exactly the same as that in the SM [34]. It is expected that the new contributions from flavor-changing $Z'$ models in which $Z'$ mediates vector and axial vector interactions would enhance the branching ratios of $B \to \pi\pi$ decay significantly. However, the $Z'$ flavor-changing couplings are subjected to constraints from relevant experimental measurements.

Considering the above effect of $Z'$ boson, the branching ratio of the $B \to \pi\pi$ decays is found to be

$$[\mathrm{B}(B \to \pi\pi)]_{Z'} = \left[\frac{1}{8\pi}\frac{|\vec{P}|}{m_B^2}|A(B \to \pi\pi)|^2 \frac{1}{\Gamma_{tot}}\right]_{Z'}. \qquad (20)$$

In the next section we use this formula for the calculation of branching ratios for $B \to \pi\pi$ decays in the presence of $Z'$ boson.

## 4. NUMERICAL RESULTS AND DISCUSSIONS

In this section, we calculate the branching ratios for $B \to \pi\pi$ decays in the presence of $Z'$ boson using all the recent data [50]: $m_{\pi^\pm} = 139.570$ MeV, $m_{\pi^0} = 134.976$ MeV, $m_{B^\pm} = (5279.0 \pm 0.5)$ MeV, $m_{B^0} = (5279.4 \pm 0.5)$ MeV, $M_Z = 91.1876$ GeV, mean lifetime $\tau_{B^\pm} = (1.671 \pm 0.018) \times 10^{-12}$ s, $\tau_{B^0} = (1.536 \pm 0.014) \times 10^{-12}$ s, Fermi constant $G_F = 1.16639 \times 10^{-5}$ GeV$^{-2}$, decay constants $f_B = 190$ MeV, $f_\pi = 130$ MeV and $\sin^2\theta_W = 0.23$. The $Z'$ parameters $\zeta_d^{LR} \sim 0.05$, $\xi_d^{LR} \sim O(10^{-2})$ and $\phi_d^L \sim -48^0$ [49]. We assume that the product $\left|B_{db}^{L*} B_{dd}^L\right|$ is numerically about the same as $\left|V_{tb} V_{td}^*\right|$. Since the $Z'$ boson has not yet been discovered, its mass is unknown. There are stringent recent limits on the mass of an extra $Z'$ boson obtained by CDF, DØ and LEP 2, and on the Z-$Z'$ mixing angle $\theta_{ZZ'}$ [51]. The precision electroweak (EW) data strongly constrain on $\theta_{ZZ'}$ to very small values, $\left|\theta_{ZZ'}\right| \leq 8.1 \times 10^{-3}$. The lower mass limits on $Z'$ bosons as obtained by CDF, DØ and LEP 2 is $M_{Z'} > 434$ GeV. However, in perturbative heterotic string models with supergravity mediated supersymmetry breaking it is found that $Z'$ mass could be less than 1 TeV [22,52,53]. In a study of B meson decays with $Z'$-mediated FCNCs [47], they study the $Z'$ boson in the mass range of a few hundred GeV to 1 TeV. Our investigations in both the left-right symmetric model [54] and potential model [55] give the mass of $Z'$ boson around 1 TeV. There are thus good motivations for an extra $Z'$ boson, with a mass range 500 GeV – 1 TeV [56].

In general, the value of $g'/g$ is undetermined [57]. However, generically, one expects that $g'/g \approx 1$ if both U(1) groups have the same origin from some grand unified theory. In order to get significant contribution due to $Z'$-mediated FCNCs, we have varied the ratio $g'/g$ from 1 to 5 in our calculations.

From equation (20), it is clear the branching ratio for $B \to \pi\pi$ decay process depends upon the value of $g'/g$ and $M_{Z'}$ [Since the amplitude $A(B \to \pi\pi)$ includes the $Z'$ contribution and calculated by using equation (16)]. We have tried to show the variation of branching ratios for the $B \to \pi\pi$ decays by varying $g'/g$ from 1 to 5 and $M_{Z'}$ from 500 GeV to 1 TeV. The results are shown in Tables 1–5 respectively. We find that the values of branching ratios are larger than the SM value as well as the recently observed experimental values. Our calculated branching ratios $B(B^0 \to \pi^0\pi^0)$ are enhanced significantly from their SM value. For $g'/g = 1$, branching ratios $B(B^0 \to \pi^0\pi^0)$ are very close to the recently observed experimental values [Eq. (9)] and for other values of $g'/g$ branching ratios are more. Our calculated branching ratios $B(B^0 \to \pi^+\pi^-)$ and $B(B^+ \to \pi^+\pi^0)$ are also enhanced from the SM value as well as the recently observed experimental values. We can also argue that there have been significant enhancements in the branching ratios for a lighter $Z'$ boson. In the experimental side, these results may be inaccessible at presently running B factories. However, it is large enough for LHCb and/or Super B factories.

From our analysis, we conclude that the branching ratios for $B \to \pi\pi$ decay processes are enhanced from its standard model value. This is due to the effect of $Z'$-mediated flavor-changing neutral currents on the $B \to \pi\pi$ decays. Hence, there may be the possibility of new physics effects in the $B \to \pi\pi$ decays [4, 58]. Furthermore, future observations of these decays would help us to constrain the mass of $Z'$ boson within the model. These facts lead to enrichment in the phenomenology of both the $Z'$-mediated FCNCs and $B \to \pi\pi$ decays.

**Table 1.** Branching ratios of the $B \to \pi\pi$ decays for $g'/g = 1$.

| $M_{Z'}$ (GeV) | $g'/g$ | $B(B^0 \to \pi^+ \pi^-)$ $(10^{-6})$ | $B(B^0 \to \pi^0 \pi^0)$ $(10^{-6})$ | $B(B^+ \to \pi^+ \pi^0)$ $(10^{-6})$ |
|---|---|---|---|---|
| 500 | 1 | 9.40 ± 0.10 | 1.24 ± 0.01 | 6.35 ± 0.13 |
| 600 | 1 | 9.30 ± 0.01 | 1.24 ± 0.004 | 6.20 ± 0.08 |
| 700 | 1 | 9.25 ± 0.03 | 1.24 ± 0.002 | 6.15 ± 0.05 |
| 800 | 1 | 9.10 ± 0.03 | 1.20 ± 0.007 | 6.10 ± 0.05 |
| 900 | 1 | 8.90 ± 0.18 | 1.20 ± 0.006 | 6.00 ± 0.12 |
| 1000 | 1 | 8.90 ± 0.14 | 1.20 ± 0.005 | 6.00 ± 0.10 |

**Table 2.** Branching ratios of the $B \to \pi\pi$ decays for $g'/g = 2$.

| $M_{Z'}$ (GeV) | $g'/g$ | $B(B^0 \to \pi^+ \pi^-)$ $(10^{-6})$ | $B(B^0 \to \pi^0 \pi^0)$ $(10^{-6})$ | $B(B^+ \to \pi^+ \pi^0)$ $(10^{-6})$ |
|---|---|---|---|---|
| 500 | 2 | 11.43 ± 0.64 | 1.56 ± 0.21 | 7.70 ± 0.28 |
| 600 | 2 | 10.62 ± 0.43 | 1.44 ± 0.22 | 7.16 ± 0.20 |
| 700 | 2 | 10.15 ± 0.12 | 1.36 ± 0.20 | 6.84 ± 0.16 |
| 800 | 2 | 9.85 ± 0.66 | 1.32 ± 0.22 | 6.64 ± 0.33 |
| 900 | 2 | 9.65 ± 0.18 | 1.28 ± 0.21 | 6.50 ± 0.60 |
| 1000 | 2 | 9.50 ± 0.60 | 1.28 ± 0.20 | 6.40 ± 0.61 |

**Table 3.** Branching ratios of the $B \to \pi\pi$ decays for $g'/g = 3$.

| $M_{Z'}$ (GeV) | $g'/g$ | $B(B^0 \to \pi^+ \pi^-)$ $(10^{-6})$ | $B(B^0 \to \pi^0 \pi^0)$ $(10^{-6})$ | $B(B^+ \to \pi^+ \pi^0)$ $(10^{-6})$ |
|---|---|---|---|---|
| 500 | 3 | $15.03 \pm 1.01$ | $2.04 \pm 0.33$ | $10.13 \pm 1.0$ |
| 600 | 3 | $12.99 \pm 0.87$ | $1.72 \pm 0.29$ | $8.75 \pm 0.84$ |
| 700 | 3 | $11.83 \pm 0.79$ | $1.56 \pm 0.27$ | $8.33 \pm 0.78$ |
| 800 | 3 | $11.10 \pm 0.74$ | $1.48 \pm 0.25$ | $7.49 \pm 0.72$ |
| 900 | 3 | $10.62 \pm 0.71$ | $1.40 \pm 0.23$ | $7.15 \pm 0.70$ |
| 1000 | 3 | $10.28 \pm 0.69$ | $1.36 \pm 0.22$ | $6.93 \pm 0.66$ |

**Table 4.** Branching ratios of the $B \to \pi\pi$ decays for $g'/g = 4$.

| $M_{Z'}$ (GeV) | $g'/g$ | $B(B^0 \to \pi^+ \pi^-)$ $(10^{-6})$ | $B(B^0 \to \pi^0 \pi^0)$ $(10^{-6})$ | $B(B^+ \to \pi^+ \pi^0)$ $(10^{-6})$ |
|---|---|---|---|---|
| 500 | 4 | $20.89 \pm 1.40$ | $2.80 \pm 0.47$ | $14.08 \pm 1.20$ |
| 600 | 4 | $16.69 \pm 1.12$ | $2.24 \pm 0.37$ | $11.25 \pm 1.0$ |
| 700 | 4 | $14.39 \pm 0.97$ | $1.92 \pm 0.32$ | $9.70 \pm 0.90$ |
| 800 | 4 | $12.98 \pm 0.87$ | $1.72 \pm 0.29$ | $8.75 \pm 0.80$ |
| 900 | 4 | $12.06 \pm 0.69$ | $1.40 \pm 0.23$ | $7.98 \pm 0.66$ |
| 1000 | 4 | $11.42 \pm 0.68$ | $1.36 \pm 0.22$ | $6.99 \pm 0.60$ |

**Table 5.** Branching ratios of the $B \to \pi\pi$ decays for $g'/g = 5$.

| $M_{Z'}$ (GeV) | $g'/g$ | $B(B^0 \to \pi^+ \pi^-)$ $(10^{-6})$ | $B(B^0 \to \pi^0 \pi^0)$ $(10^{-6})$ | $B(B^+ \to \pi^+ \pi^0)$ $(10^{-6})$ |
|---|---|---|---|---|
| 500 | 5 | $29.85 \pm 2.01$ | $4.04 \pm 0.67$ | $20.13 \pm 2.0$ |
| 600 | 5 | $22.14 \pm 1.50$ | $3.00 \pm 0.50$ | $14.93 \pm 1.4$ |
| 700 | 5 | $18.05 \pm 1.22$ | $2.44 \pm 0.40$ | $12.17 \pm 1.0$ |
| 800 | 5 | $15.62 \pm 0.87$ | $2.08 \pm 0.35$ | $10.53 \pm 1.0$ |
| 900 | 5 | $14.05 \pm 0.95$ | $1.88 \pm 0.31$ | $9.48 \pm 0.90$ |
| 1000 | 5 | $12.98 \pm 0.88$ | $1.72 \pm 0.29$ | $8.75 \pm 0.80$ |

**REFERENCES**

1. R. Fleischer, hep-ph/0701217; A. J. Bevan, hep-ex/0701031; H-n. Li, Pramana - J. Phys. **67,** 755 (2006) [hep-ph/0605331]; Y. Yang, F. Su, G. Lu and H. Hao, Eur. Phys. J. C **44,** 243 (2005).
2. A. J. Buras, R. Fleischer, S. Recksiegel and F. Schwab, Phys. Rev. Lett. **92,** 101804 (2004) [hep-ph/0312259]; Nucl. Phys. B **697,** 133 (2004) [hep-ph/0402112]; R. Fleischer, hep-ph/0505018; R. Fleischer, Int. J. Mod. Phys. A, **21** 664 (2006).
3. A. J. Buras, R. Fleischer, S. Recksiegel and F. Schwab, Eur. Phys. J. C **45,** 701 (2006); Acta Phys. Polon. B **36,** 2015 (2005); R. Arnowitt, B. Dutta, B. Hu and S. Oh Phys. Lett. B **633,** 748 (2006); X-G. He and B. H. J. McKellar, hep-ph/0410098; Y-Y. Charng and H-n. Li, Phys. Rev. D **71,** 014036 (2005) [hep-ph/0410005].
4. S. Baek, J. High Energy Phys. **0607,** 025 (2006) [hep-ph/0605094]; S. Baek, P. Hamel, D. London, A. Datta and D. A. Suprun, Phys. Rev. D **71,** 057502 (2005) [hep-ph/0412086]; Y.-L. Wu and Y.-F. Zhou, Phys. Rev. D **72,** 034037 (2005) [hep-ph/0503077]; W. T. Ford, hep-ex/0510058
5. Heavy Flavour Averaging Group, http://www.slac.stanford.edu/xorg/hfag/ [Updated result].
6. Belle Collaboration, http://belle.kek.jp/
7. K. Abe *et al.* [Belle Collaboration], hep-ex/0608033; hep-ex/0608049; hep-ex/0609015; J. Schumann [Belle Collaboration], Phys. Rev. Lett. **97,** 061802 (2006).
8. BaBar Collaboration, http://www-public.slac.stanford.edu/babar/Babar Publications.
9. B. Aubert *et al.* [BaBar Collaboration], hep-ex/0607063; hep-ex/0607096; hep-ex/0607106; hep-ex/0608003; hep-ex/0608036; Phys. Rev. Lett. **94,** 191802 (2005) [hep-ex/0502017]; Phys. Rev. Lett. **95,** 131803 (2005) [hep-ex/0503035]; Phys. Rev. D **73,** 071102 (2006) [hep-ex/0603013].
10. E. Kou and T. N. Pham, Phys. Rev. D **74,** 014010 (2006) [hep-ph/0601272]; T. N. Pham hep-ph/0610063.
11. Y-L. Wu, Y-F. Zhou and C. Zhuang, hep-ph/0609006; S. Mishima, hep-ph/0605226.
12. J-F. Cheng, Y-N. Gao, C-S. Huang and H-X. Wu, Phys. Lett. B **637,** 260 (2006); G. Buchalla and A. S. Safir, Eur. Phys. J. C **45,** 109 (2006).
13. M. Imbeault *et al.*, hep-ph/0608169; K. Abe *et al.*, hep-ex/0608035.
14. S. Nandi and A. Kundu, hep-ph/0407061.
15. M. Wirbel, B. Stech and M. Bauer, Zeit. Phys. C **29,** 637 (1985); M. Bauer, B. Stech and M. Wirbel, Zeit. Phys. C **34,** 103 (1987); A. Ali, G. Kramer and C-D. Lu, Phys. Rev. D **58,** 094009 (1998); Phys. Rev. D **59,** 014005 (1999).
16. M Benke *et al.*, Phys. Rev. Lett. **83,** 1914 (1999) [hep-ph/9905312]; Nucl. Phys. B **591,** 313 (2000) [hep-ph/0006124]; Nucl. Phys. B **606,** 245 (2001)[hep-ph/0104110].
17. Y-Y. Keum, H-n. Li and A. I. Sanda, Phys. Lett. B **504,** 6 (2001) [hep-ph/0004004]; Phys. Rev. D **63,** 054008 (2001) [hep-ph/0004173]; Y-Y. Keum and H-n. Li, Phys. Rev. D **63,** 074006 (2001) [hep-ph/0006001]; C-D. Lu, K. Ukai and M. Z. Yang Phys. Rev. D **63,** 074009 (2001) [hep-ph/0004213]; Y-Y. Keum and A. I. Sanda, Phys. Rev. D **67,** 054009 (2003); H-n. Li, S. Mishima and A. I. Sanda, Phys. Rev. D **72,** 114005 (2005) [hep-ph/0508041].

18. C. W. Bauer, I. Z. Rothstein and I. W. Stewart, hep-ph/0510241; C. W. Bauer *et al.*, hep-ph/0401188 and references therein.
19. R. Mohanta, hep-ph/0503225; A. K. Giri and R. Mohanta, Eur. Phys. J. C, **45,** 151 (2006); J. A. Aguilar-Saavedra, Acta Phys. Polon. B **35,** 2695 (2004).
20. J. D. Lykken, hep-ph/9610218; P. Langacker, hep-ph/0308033.
21. S. Choudhuri, S. W. Chung, G. Hockney and J. Lykken, Nucl. Phys. B **456,** 89 (1995) [hep-ph/9501361]. M. Masip and A. Pamarol, Phys. Rev. D **60,** 096005 (1999) [hep-ph/9902467].
22. J. Hewett and T. Rizzo, Phys. Rep. **183,** 193 (1989).
23. F. Abe *et al.* [CDF Collaboration], Phys. Rev. Lett. **77,** 438 (1996); Phys. Rev. Lett **79,** 2192 (1997).
24. J. Erler and P. Langacker, Phys. Lett. B **456,** 68 (1999) [hep-ph/9903476]; J. Erler and P. Langacker, Phys. Rev. Lett. **84,** 212 (2000) [hep-ph/9910315]; The LEP Electroweak Working Group and SLD Heavy Flavour Group, hep-ex/0212036.
25. P. Langacker and M. Plumacher, Phys. Rev. D **62,** 013006 (2000) [hep-ph/0001204].
26. S. Sahoo, Indian J. Phys. **80** (2), 191 (2006); C. T. Hill, Phys. Lett. B **345,** 483 (1995); R. S. Chivukula, E. H. Simmons, J. Terning, Phys. Lett. B **331,** 383 (1994); H. Georgi, E. E. Jenkis, E. H. Simmons, Phys. Rev. Lett. **62,** 2789 (1989); H. Georgi, E. E. Jenkis, E. H. Simmons, Nucl. Phys. B **331,** 541 (1990).
27. A. Leike, Phys. Rep. **317,** 143 (1999) [hep-ph/9805494].
28. J. H. Jeon, C. S. Kim, J. Lee and C. Yu, Phys. Lett. B **636,** 270 (2006); X-G, He and G. Valencia, hep-ph/0605202; D. Feldman, Z. Liu and P. Nath, Phys. Rev. Lett. **97** 021801 (2006) [hep-ph/0603039]; B. B. Deo and L. Maharana, Phys. Lett. B **461,** 105 (1999); K. Cheung, C – W. Chiang, N. G. Deshpande and J. Jiang, hep-ph/0604223.
29. K. S. Babu, C. Kolda and J. March-Russell, Phys. Rev. D **54,** 4635 (1996) [hep-ph/9603212]; Phys. Rev. D **57,** 6788 (1998) [hep-ph/9710441]; T. G. Rizzo, Phys. Rev. D **59,** 015020 (1998) [hep-ph/9806397]; F. del Aguila, J. Moreno and M. Quiros, Nucl. Phys. B **372,** 3 (1992); **361,** 45 (1991).
30. S. Sahoo and L. Maharana, Phys. Rev. D **69,** 115012 (2004); V. Berger, C-W. Chiang, J. Jiang and P. Langacker, Phys. Lett. B **596,** 229 (2004) [hep-ph/0405108]; K. Leroux and D. London, Phys. Lett. B **526,** 97 (2002) [hep-ph/0111246]; M. Gronau and D. London, Phys. Rev. D **55,** 2845 (1997); E. Nardi and Silverman, Phys. Rev. D **48,** 1240 (1993); J. Bernabeu, E. Nardi and D. Tommasini, *Nucl. Phys.* B **409,** 69 (1993); [hep-ph/9306251]; S. Baek, J. H. Jeon and C. S. Kim, hep-ph/0607113.
31. A. Arhrib, K. Chung, C-W. Chiang and T.- C. Yuan, hep-ph/0602175.
32. J-F. Cheng, Y-N. Gao, C-S. Huang and H-X. Wu, hep-ph/0612116.
33. C-D. Lu, K. Ukai and M-Z. Yang, Phys. Rev. D **63,** 074009 (2001).
34. G. Buchalla, A. J. Buras and M. E. Lautenbacher, Rev. Mod. Phys. **68,** 1125 (1996) [hep-ph/9512380]; A. J. Buras, hep-ph/9806471.
35. A. Ali and C. Greub, Phys. Rev. D **57,** 2996 (1998) [hep-ph/ 9707251]; Y. H. Chen, H. Y. Cheng, B. Tseng and K. C. Yang, Phys. Rev. D **60,** 094014 (1999); H. Y. Cheng, and K. C. Yang, Phys. Rev. D **62,** 054029 (2000); G. Altarelli, G. Curci, G. Martinelli and S. Petrarca, Nucl. Phys. B **187,** 461 (1981); M. Beneke, Th Fledmann, and D. Seidel, Nucl. Phys. B **612,** 25 (2001).
36. S. Khalil, Phys. Rev. D **72,** 035007 (2005) [hep-ph/0505151].

37. Y. Grossman, M. Neubert and A. L. Kagan, hep-ph/9909297.
38. M. Gronau, O. F. Hernandez, D. London and J. L. Rosner, Phys. Rev. D **50,** 4529 (1994) [hep-ph/9404283]; Phys. Rev. D **52,** 6356 (1995); Phys. Rev. D **52,** 6374 (1995).
39. C-W. Chiang and Y-F. Zhou, hep-ph/0609128.
40. M. Beneke and M. Neubert, Nucl. Phys. B **675,** 333 (2003) [hep-ph/0308039].
41. Y. D. Yang, R-M. Wang and G. R. Lu, Phys. Rev. D **73,** 015003 (2006) [hep-ph/0509273].
42. L. Wolfenstein, Phys. Rev. Lett. **51,** 1945 (1983).
43. Y. Y. Chang and H-n. Li, Phys. Lett. B **594,** 185 (2004); S. Mishima and T. Yoshikawa, Phys. Rev. D **70,** 094024 (2004) [hep-ph/0408090]; T. Yoshikawa, Phys. Rev. D **68,** 054023 (2003).
44. M. Beneke and D. Yang, Nucl. Phys. B **736,** 34 (2006).
45. A. Deandrea *et al.*, Int. J. Mod. Phys. A **21,** 4425 (2006) [hep-ph/0508083].
46. S. Baek, F. J. Botella, D. London and J. P. Silva, Phys. Rev. D **72,** 114007 (2005) [hep-ph/0509322]; Phys. Rev. D **72,** 036004 (2005) [hep-ph/0506075].
47. V. Berger, C-W. Chiang, P. Langacker and H. S. Lee, Phys. Lett. B **580,** 186 (2004) [hep-ph/0310073].
48. V. Berger, C-W. Chiang, P. Langacker and H. S. Lee, Phys. Lett. B **598,** 218 (2004) [hep-ph/0406126].
49. Q. Chang, X-Q. Li and Y.-D. Yang, arXiv:1003.6051 v1 [hep-ph] 31 Mar 2010.
50. C. Amsler *et al.* [Particle Data Group], Phys. Lett. B **667,** 1 (2008).
51. J. Erler, P. Langacker, S. Munir and E. Rojas, J. High Energy Phys. **0908,** 017 (2009) [arXiv: 0906.2435 [hep-ph]]; arXiv: 0910.0269 [hep-ph].
52. M. Cvetic and S. Godfrey, hep-ph/9504216; S. Godfrey, hep-ph/0201092; hep-ph/0201093
53. M. Cvetic and P. Langacker, Mod. Phys. Lett. A **11,** 1247 (1996); M. Cvetic, D. A. Demir, J. R. Espinosa, L. Everett and P. Langacker, Phys. Rev. D **58,** 119905 (1998).
54. S. Sahoo, L. Maharana, S. Acharya and A. Roul, Int. J. Mod. Phys. A **20** 2625 (2005).
55. S. Sahoo, A. K. Meikap and L. Maharana, Mod. Phys. Lett. A **21,** 275 (2006).
56. A. Cordero-Cid, G. Tavares-Velasco and J. J. Toscano, hep-ph/0507135.
57. M. Cvetic and B. W. Lynn, Phys. Rev. D **35,** 51 (1987)
58. R. Fleischer, S. Recksiegel and F. Schwab, hep-ph/0702275.